\documentclass{optica-article}

\journal{opticajournal} 

\articletype{Research Article}
\pdfoutput=1

\usepackage{graphicx}
\usepackage{amsmath}
\usepackage{mathtools}
\usepackage{upgreek}
\usepackage{bm}
\usepackage{geometry}
\usepackage{multirow,booktabs}
\usepackage{threeparttable}

\begin{document}

\title{Weakly supervised learning for pattern classification in serial femtosecond crystallography}

\author{Jianan Xie\authormark{1,3,8}, Ji Liu\authormark{2,8}, Chi Zhang\authormark{1,3}, Xihui Chen\authormark{3,4}, Ping Huai\authormark{5,6}, Jie Zheng\authormark{2,7}, Xiaofeng Zhang\authormark{5,*}}

\address{\authormark{1}School of Physical Science and Technology, ShanghaiTech University, 393 Middle Huaxia Road, Shanghai 201210, China\\
	\authormark{2}School of Information Science and Technology, ShanghaiTech University, Shanghai 201210, China\\
	\authormark{3}Shanghai Advanced Research Institute, Chinese Academy of Sciences, 99 Haike Road, Shanghai 201210, China\\
	\authormark{4}University of Chinese Academy of Sciences, Beijing 100049, China\\
	\authormark{5}Center for Transformative Science, ShanghaiTech University, 393 Middle Huaxia Road, Shanghai 201210, China\\
	\authormark{6}ShanghaiTech-SARI Joint Lab for Photon Science, Shanghai Advanced Research Institute, Chinese Academy of Sciences, 99 Haike Road, Shanghai 201210, China\\
	\authormark{7}Shanghai Engineering Research Center of Intelligent Vision and Imaging, Shanghai 201210, China\\
	\authormark{8}The authors contributed equally to this work.}

\email{\authormark{*}zhangxf2@shanghaitech.edu.cn}


\begin{abstract*}
Serial femtosecond crystallography at X-ray free electron laser facilities opens a new era for the determination of crystal structure. However, the data processing of those experiments is facing unprecedented challenge, because the total number of diffraction patterns needed to determinate a high-resolution structure is huge. Machine learning methods are very likely to play important roles in dealing with such a large volume of data. Convolutional neural networks have made a great success in the field of pattern classification, however, training of the networks need very large datasets with labels. This heavy dependence on labeled datasets will seriously restrict the application of networks, because it is very costly to annotate a large number of diffraction patterns. In this article we present our job on the classification of diffraction pattern by weakly supervised algorithms, with the aim of reducing as much as possible the size of the labeled dataset required for training. Our result shows that weakly supervised methods can significantly reduce the need for the number of labeled patterns while achieving comparable accuracy to fully supervised methods.
\end{abstract*}

\section{Introduction}
\begin{figure}[ht!]
	\centering
	\includegraphics[width=0.95\textwidth]{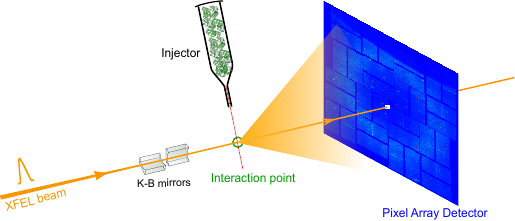}
	\caption{Diagram of XFEL SFX experimental setup. X-ray pulses hit the samples and then produce diffraction patterns on the pixel array detector. Microcrystals are continuously injected into XFEL beam.}
	\label{fig:sfx-diagram}
\end{figure}

X-ray crystallography at synchrotron radiation light sources plays an important role in the determination of macromolecular structure, but radiation damage has always been a difficult problem in the measurement. The scientists had to increase crystal sizes and put the crystals under cryo-cooled conditions to allow for higher radiation tolerances and thus improve the structural resolution. The development of X-ray free electron laser (XFEL) makes it possible to produce X-ray pulse with extreme peak brilliance and ultrashort pulse width, thereby enables the radiation damage to be overcome by ``diffraction-before-destruction'' principle \cite{Chapman2006,Chapman2014}. In contrast to traditional X-ray crystallography, serial femtosecond crystallography (SFX) at XFELs is able to measure atomic structure of microcrystals in room temperature \cite{Chapman2011,Johansson2017,Martin-Garcia2016}. XFEL pulse is so intense that the sample is destroyed after interacting with the pulse, and therefore in SFX experiments samples must be continuously replenished by the delivery system (see Fig. \ref{fig:sfx-diagram}). X-ray pulses interact with randomly oriented crystals and then generate diffraction frames on a pixel array detector, but actually the ratio of XFEL pulses hitting the samples is very small. For example,  only 6.1\% patterns were identified as effective frames in the measurement of Photosystem I \cite{Chapman2011} at the Linac Coherent Light Source (LCLS)~\cite{White2015}, and only 3.4\% frames were found to contain crystal diffraction during the HEWL data acquisition \cite{Wiedorn2018} at the European XFEL \cite{Euxfel2006}. In order to determine the structure of microcrystals at atomic scale, it is usually necessary to collect millions of diffraction frames. In the study of native nanocrystalline granulovirus, a total of 1.5 millions of collected detector frames yielded 2 \AA~resolution, and the analysis clearly showed that averaging more diffraction frames improved all figures of merit with a better signal-to-noise ratio\cite{Gati2017}.

Considering the poor hit rate, most collected frames contain only noise and thus are useless in SFX experiments. Consequently, selecting patterns with crystal diffraction from the tremendous volume of raw frames is an essential step in the processing of SFX data. Several open-source software packages have been developed for automatic hit-finding and pattern classification, e.g., $ Cheetah $~\cite{Barty2014} and \textit{DIALS}~\cite{Winter2018}. In practice, the parameters of those toolkits usually need manual tuning due to some complex factors in experiments such as low signal-to-noise ratio, detector artifact, unstable beam etc. Machine learning algorithms provide another approach to classifying the millions of diffraction frames, especially the convolutional neural networks (CNNs), which have achieved remarkable success in the field of pattern recognition. Compared with traditional programs, CNN can encode both human expertise and latent characters of the figures, and is less sensitive to some noise. In recent years several works have demonstrated the performance of CNNs in the classification of diffraction patterns~\cite{Ke2018,Zimmermann2019,Shi2019}.

However, CNN is a typical supervised algorithm, which means that a large dataset with labels is needed to train the network. This heavy dependency on large size of labeled dataset has hindered the application of CNN, because it is very costly to annotate millions of figures in each experiment, especially for diffraction patterns with complex features. In contrast to supervised learning, training a model by unlabeled datasets is called unsupervised learning. Unfortunately, so far the performance of unsupervised algorithms is not so good in the classification of diffraction patterns. Recently, a kind of methods called weakly supervised learning has been proposed and developed rapidly~\cite{Zhou2018}, which managed to train models with the combination of small labeled datasets and large unlabeled datasets. Because it is very easy to get unlabeled diffraction patterns, weakly supervised algorithms are promising solutions to reducing the heavy dependence of CNNs on human annotations, thereby promoting the application of CNNs in the processing of SFX data. In this paper we present our study of using three weakly supervised CNN models to identify SFX frames with crystal diffraction.

\section{Methods}\label{sec:methods}
\subsection{Train, validate and test}
The training of a CNN model consists of multiple epochs, and each epoch typically includes three steps: training, validation ant test. Accordingly, the whole dataset is usually divided into three parts: training set, validation set and test set. In the training steps the model learns features of the input data and updates its weights. The task of validation steps is to monitor the performance of the model with the goal of better control over the training process. The generalization performance of a model should be evaluated in test steps before applying it to new data, and the dataset used in this stage must not be fed to the model in the training or validation steps. Test steps are very important to avoid overfitting, which means the models perform extremely well on the training set but poorly on the test set. Overfitting is a very common problem in machine learning, which can be overcome by reducing the complexity of the model or increasing the size of the training set.

\subsection{Convolutional neural network}
The basic architecture of a neural network can be divided into a number of fully connected (FC) layers and activation layers. Except for the last layer, which is often called the output layer, the output of an FC layer is passed to an activation layer, and the output of an activation layer is fed to the next FC layer. The FC layer acts as a linear mapping, while the activation layer plays the role of nonlinear mapping. A model consists of many FC layers is very time consuming to be trained, because the number of trainable parameters is large. In CNNs most FC layers are replaced by convolutional layers, thus greatly reducing the number of parameters to be trained. The most remarkable difference between an FC layer and a convolutional layer is that each neuron in the FC layer has its own weights, while neurons in some region of the convolutional layer share the same weights. In addition to weights sharing, pooling layers are also used in CNN to downsample the outputs of previous layers and thus to further reduce the number of trainable parameters. Compared with the FC networks, CNNs not only can be trained more efficiently but also are more powerful in the field of pattern recognition.

\begin{figure}[ht!]
	\centering
	\includegraphics[width=0.95\textwidth]{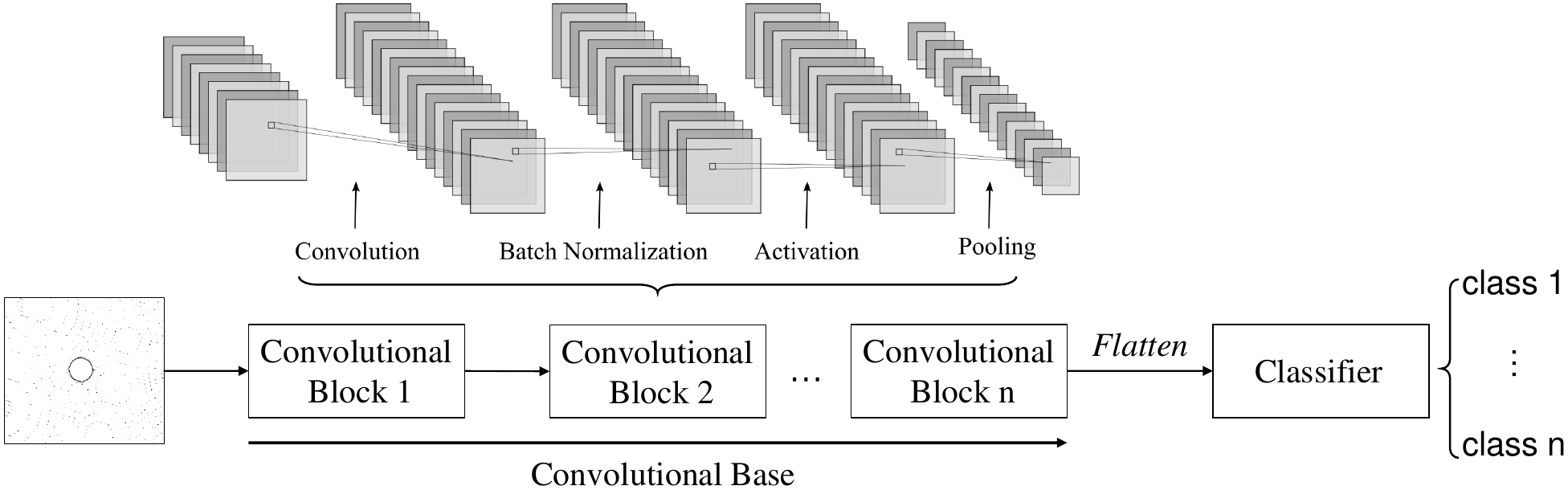}
	\caption{A typical CNN can be roughly divided into a convolutional base and a classifier. The convolutional base extracts features of the input data and then passes them to the classifier. The output of classifier is a 1D vector, each element of which represents the predicted probability of a class. The convolutional base consists of several convolutional blocks, and each block is built by stacking a convolution layer, a batch normalization layer, an activation layer, and a pooling layer. It is common to add a dropout layer in each of the top two or three blocks to prevent overfitting.}
	\label{fig-typical-cnn}
\end{figure}

A typical CNN for classification task can be roughly divided into two parts: a convolutional base and a classifier (Fig. \ref{fig-typical-cnn}). The convolutional base extracts and encodes the latent features of the input data, and then the encoded features are transformed into a one-dimensional (1D) vector by a flatten layer so that it can be fed to the classifier.
The output of the classifier is a 1D vector commonly interpreted as approximating a probability distribution, with each element representing the predicted probability of a class. Convolutional base consists of several blocks, each of which is usually constructed by stacking a convolutional layer, a batch normalization (BN) layer \cite{Ioffe2015} , an activation layer, and a pooling layer. In the training steps, the BN layer zero-centers and normalizes the output of the convolutional layer, which has been proved to be significant for improving the training process. A classifier usually consists of two or three FC layers and their activation layers. Dropout is widely used to suppress the overfitting of CNNs, which actually deactivates part of neurons randomly \cite{Hinton2012, Agarwal2014}. Adding one or a few dropout layers on the top of convolutional base has been proved to be a simple but efficient strategy to prevent overfitting.

A CNN is usually trained through the backpropagation (BP) algorithm \cite{Rumelhar1987}, which divides a train step into two processes: First, the response of each layer is propagated forward, i.e., from the input layer to the output layer, and the loss representing the difference between the output and the label is calculated by a predefined function; the second process starts with the loss value and works backward from the output layer to the input layer, updating the weights of each layer in the direction of decreasing the loss value:
\begin{equation}
	\mathbf{w} \leftarrow \mathbf{w} - \eta \cdot \nabla_{\mathbf{w}} L,
\end{equation}
where $\mathbf{w}$ is the weight vector, $\eta$ is the learning rate and $L$ is the loss.

\subsection{Weakly supervised learning}
Large datasets with labels are essential for supervised learning, otherwise only overfitting or weak models can be obtained. Generally it is easy to get large datasets but costly to label them, especially for XFEL diffraction patterns since labeling them requires expertise. This severe dependency is a big obstacle for processing XFEL diffraction patterns with supervised learning models. Since it is difficult to perform classification by fully unsupervised learning, weakly supervised learning should be a better option, which tries to reduce the size of the required labeled dataset as much as possible. Particularly, training a model by the combination of a small labeled dataset and a large unlabeled dataset is a promising solution. In this section we give a brief introduction to several methods to implement weakly supervised learning.

\subsubsection{Transfer learning}
Essentially, the features learned by a CNN are encoded in its convolutional base. It is reasonable to assume that similar datasets should have similar characteristics in the latent space, thus reusing and fine tuning the base of a CNN model is a good solution to the problems with only a small labeled dataset. This method is called transfer learning \cite{Aurelien2022}, and it is probably the most popular approach when having a reusable CNN model.

\begin{figure}[ht!]
	\centering
	\includegraphics[width=0.9\textwidth]{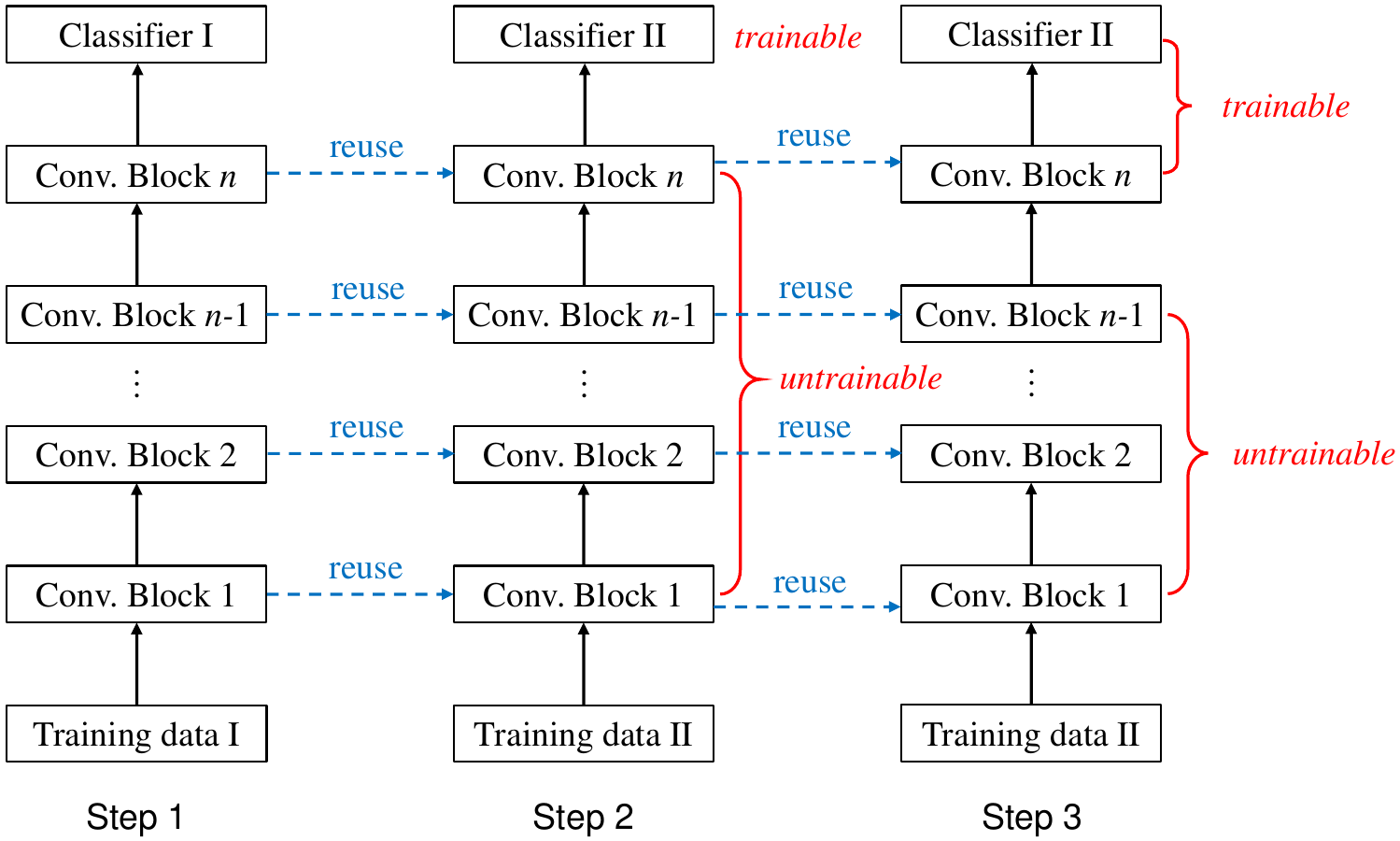}
	\caption{The proposed steps of transfer learning. The new CNN is constructed by concatenating the convolutional base of another well-trained CNN to a randomly initialized classifier. The new model can be trained with a new dataset in two steps: First, freeze the convolutional base to make its parameters unchangeable and train the model; secondly, unfreeze one or two top convolutional blocks and retrain it.}
	\label{fig-TF-learning}
\end{figure}

A three-step pipeline of transfer learning is shown in Fig. \ref{fig-TF-learning}. A new CNN can be constructed by concatenating the convolutional base of a well-trained CNN to a randomly initialized classifier, and then be trained with a new dataset in two steps: First, freeze the convolutional base to make its parameters unchangeable and train the CNN; secondly, unfreeze one or two top convolutional blocks and retrain it. The required size of the labeled dataset is considerably smaller than training a new CNN from scratch. It should be noted that the performance of transfer learning relies heavily on the similarity between the two datasets. For the diffraction data produced in SFX experiments, the similarity may be affected by several factors such as detectors, experimental methods, and samples.

\subsubsection{Dimensionality reduction and feature engineering}
Usually, only some features in the original data are useful for classification whereas the other features are useless. Learning the redundant features in a dataset often requires more complex models, which need not only longer training time, but also larger size of labeled datasets. Therefore removing the redundant features through dimensionality reduction is widely used in machine learning \cite{Reddy2020}. It transforms data from a high-dimensional space to a low-dimensional space, making the intrinsic characteristic more evident and thus reducing the demand for labels.

Another method to resolve redundant features is feature engineering \cite{Sinan2018}, where the goal is to create some new features from the dataset, which usually have higher signal-to-noise ratio and are easier to be learned by models, thereby also relieving the requirement for labels. In practice, creating new features usually reduces the dimensionality of the original data as well.

\subsubsection{Domain Adversarial neural network}\label{sec:DANN}
\begin{figure}[ht!]
	\centering
	\includegraphics[width=0.99\textwidth]{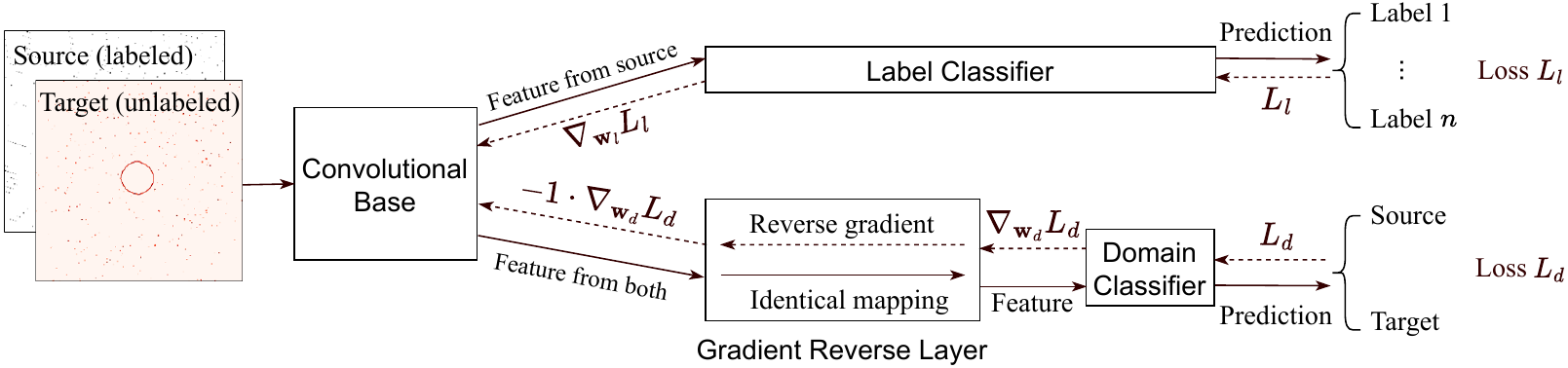}
	\caption{The structure diagram of DANN \cite{Ganin2016}. There is one convolutional base but two classifiers in this network. The solid arrows represent the forward propagation, while dashed arrows represent the backpropagation of the losses or their gradients. $L_{l}$ and $L_{d}$ denote the losses of label prediction and domain prediction respectively; ${\bf{w}}_{l}$ and ${\bf{w}}_{d}$ denote the weight vectors of label classifier and domain classifier respectively. The domain classifier is trained to distinguish the features between source and target and thus to decrease $L_{d}$, however the convolutional base is trained to increase $L_{d}$ because the backpropagated gradient is reversed by the gradient reverse layer. The adversary between domain classifier and convolutional base forces the latter to learn features from the common characteristic spaces of source and target.}
	\label{fig-DANN-architecture}
\end{figure}

The core idea of transfer learning is that two datasets collected with similar experimental setups should have similar latent characteristic spaces. Another natural strategy is to train a CNN model to learn the features of those two datasets simultaneously. 
That is exactly the idea of domain adversarial neural network (DANN) \cite{Ganin2016}. The structure diagram of DANN is shown in Fig. \ref{fig-DANN-architecture}, from which it can be seen that DANN contains one convolutional base but two classifiers, namely the label and domain classifiers (Accordingly there are two loss functions). The key design of DANN is the gradient reverse layer (GRL), which acts as an identity transformation in the forward propagation but changes the sign of the gradient, i.e., multiple it by -1, before passing it to the convolutional base in the backward propagation. In the training steps, both the labeled dataset (source) and unlabeled dataset (target) are fed to the network; features extracted from both are propagated to the GRL and domain classifier, whereas only the features extracted from source are passed to the label classifier.
The domain classifier is trained to distinguish the features between source and target and thus to decrease the loss, however the convolutional base is trained to increase the loss of domain classification because the backpropagated gradient is reversed by the GRL. The adversarial relationship between domain classifier and convolutional base forces the latter to learn features from the common characteristic spaces of source and target.
Consequently, after the training of DANN, the network composed of the convolutional base and label classifier should be able to predict the labels of target images.

\section{Results}\label{sec:results}
In this section we first introduce the datasets used in our work and then present the results of those weakly supervised methods described above. The predictions of each method are given in separate tables, with bold values denoting recall (See the definition in section \ref{sec-metrics}), i.e., the main metric we focus on in our study.

\subsection{Datasets}\label{sec:dataset}
\begin{figure}[ht!]
	\centering
	\includegraphics[width=\textwidth]{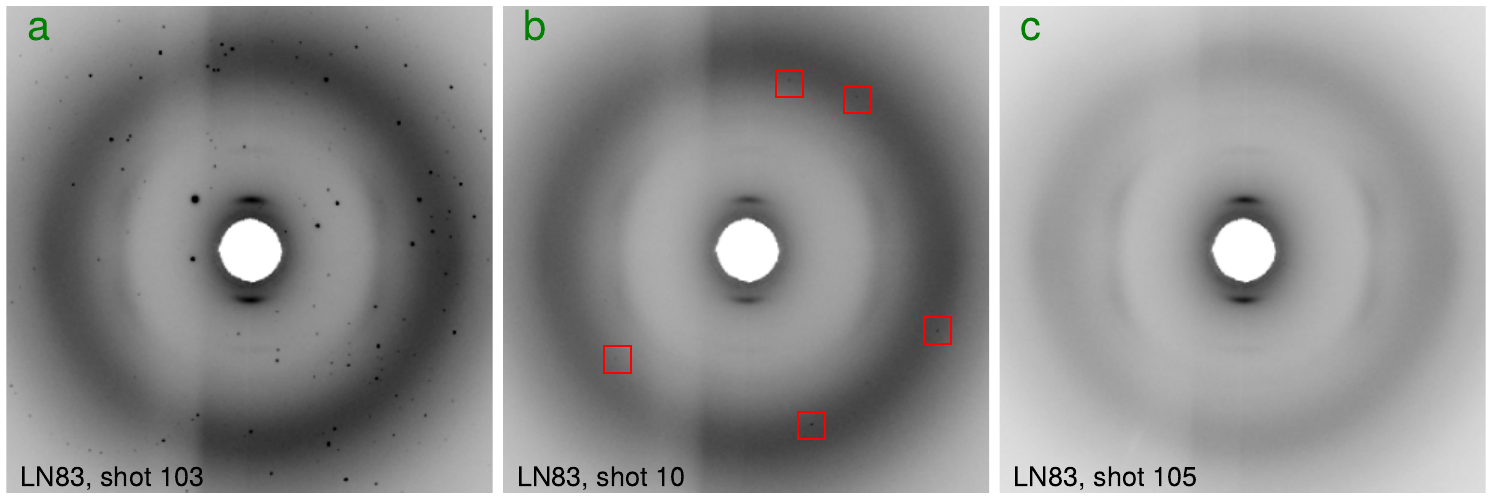}
	\caption{Examples of frames annotated as ``Hit'' (a), ``Maybe'' (b), and ``Miss'' (c). A few less visible Bragg spots are marked out by the red boxes in panel (b).}
	\label{fig:compare_hit}
\end{figure}

The datasets used in this research were downloaded from the Coherent X-ray Imaging Data Bank (CXIDB) \cite{Maia2012}, accessing number 76, at \url{http://cxidb.org/id-76.html}, which were collected at the Coherent X-ray Imaging (CXI)~\cite{Liang2015} and Macromolecular Femtosecond Crystallography (MFX)~\cite{Sierra2019} instruments of LCLS. There are five data files named L498, LG36, LN84, LN83 and LO19, each of which contains 2000 diffraction patterns. Only some of the frames contain valid diffraction signals, while others contain only background. By manually inspecting each frame, Ke \textit{et al}. \cite{Ke2018} classified all frames into three categories according to the number of Bragg spots. Frames with ten or more Bragg spots were labeled as ``Hit'', those with four to nine Bragg spots were labeled as ``Maybe'', and the rest as ``Miss''. Because there is no annotation of LG36 and the quality of images in L498 is not good, only the other three datasets are used in our work. Using the same datasets, Ke \textit{et al.} have done an excellent work in screening frames with good diffraction signature via supervised CNN models, showing that even a CNN model with simple structure outperformed the carefully tuned automatic hit-finding program \cite{Ke2018}.
Examples of three kinds of diffraction patterns are given in Fig. \ref{fig:compare_hit}, where panels a, b and c represent the patterns annotated with ``Hit'', ``Maybe'' and ``Miss'', respectively. In panel b a few less visible Bragg spots are marked out by the red boxes. Because the frames labeled as ``Maybe'' are probably useful in downstream analysis, we merge ``Hit'' and ``Maybe'' into one category in our work and thus we only need to study binary classification.

\subsubsection{Data pre-processing}
Data pre-processing on raw images is an essential step in machine learning, which usually reduces noise and makes the raw data easier to be learned by models. The pre-processing in our study is described as follows.

\begin{figure}[ht!]
	\centering
	\includegraphics[width=0.6\textwidth]{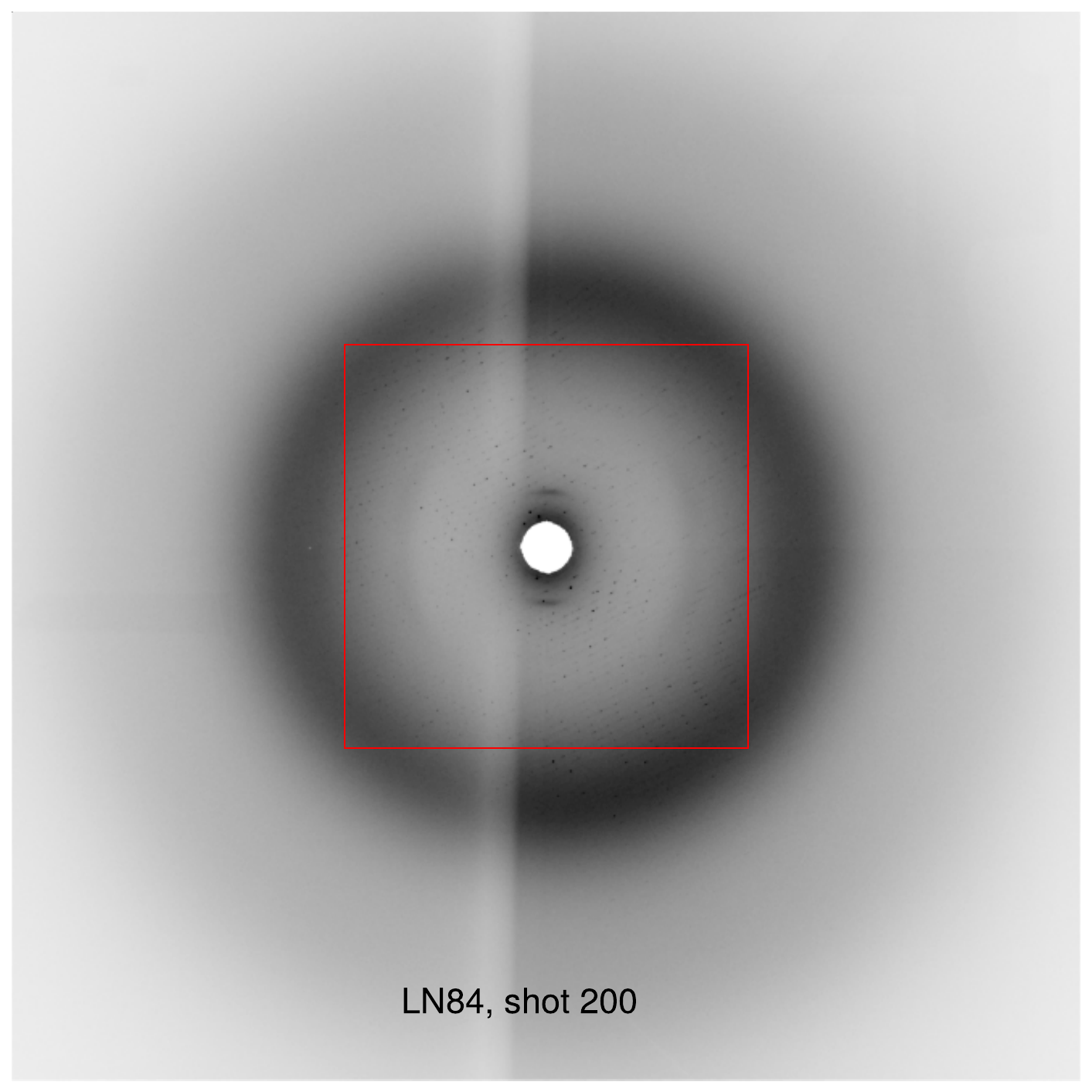}
	\caption{Example of the central cropping with a size of $724 \times 724$. Most of the Bragg spots are located in the central region.}
	\label{fig-central-crop}
\end{figure}

It is very time-consuming to train CNN models with the raw images of size $1920\times 1920$. Considering that most diffraction spots are typically located in the central region, the central cropping with a size of $724 \times 724$ is performed. This step effectively reduces the size and also preserves the as many diffraction signatures as possible (Fig. \ref{fig-central-crop}).

\begin{figure}[ht!]
	\centering
	\includegraphics[width=\textwidth]{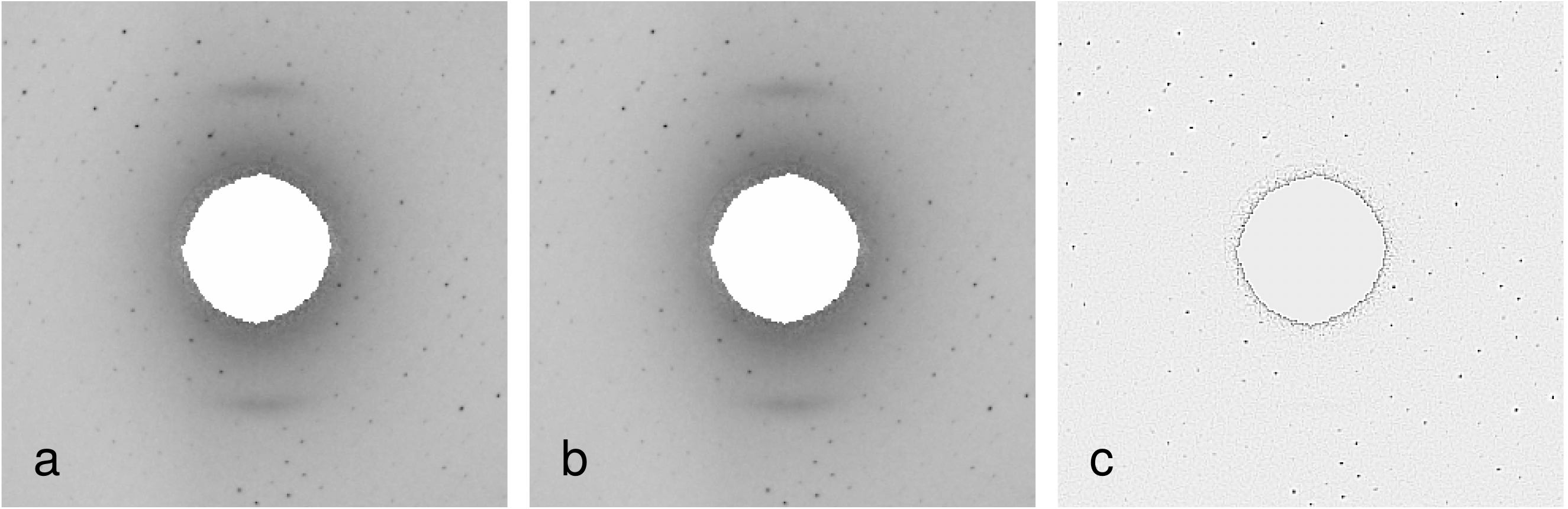}
	\caption{LN84, shot 560. The effect of contrast normalization. (a) is the experimental image without any normalization, (b) shows the same image processed by GCN, and (c) shows the same image processed by GCN and LCN.}
	\label{fig-compare-lcn}
\end{figure}

Due to the large dynamic range of the pixel array detector used in SFX experiments, the pixel values of diffraction patterns were distributed over a wide range, e.g., from 0 to 10000 or even higher. Both large pixel values and wide pixel ranges make the training of CNN models very difficult, therefore it is necessary to normalize every image so that its pixel values have a standard deviation of 1 and a mean of 0. Hence, in addition to the central cropping, we perform the same normalization as done by Ke \textit{et al.}~\cite{Ke2018}, i.e., processing each image by global contrast normalization (GCN) and local contrast normalization (LCN)~\cite{Jarrett2009}. GCN zero-centers and scales the pixel values to a small range, and then LCN removes the local background and makes the boundaries of Bragg spots more distinctive (Fig. \ref{fig-compare-lcn}). After LCN the size of each image is reduced to $720 \times 720$. The pre-processing GCN and LCN makes the CNN convergence faster during the training and also makes the trained model more robust.

\subsection{Metrics}\label{sec-metrics}
In order to evaluate the performance of a machine learning model during the training epochs, it is important to monitor some metrics in real time. The most commonly used metric in classification tasks is accuracy, i.e., the ratio of samples that are correctly classified. But in unbalanced dataset, where the class distribution is skewed, e.g., most samples belong to one or two class (majority samples) while only a small fraction of samples belong to other classes (minority samples), accuracy may be a misleading metric, because a model can still achieve a high accuracy even if it misclassifies all minority samples. In particular, using accuracy as a metric will yield very bad results if minority samples are much more important than majority samples. As mentioned above, the hit rate of an SFX experiment is very low, therefore the frames with diffraction signatures are minority samples. For unbalanced datasets, recall and precision are more useful metrics than accuracy. Usually precision and recall are dilemmatic metrics, i.e., it is difficult to improve them simultaneously, and which one is more important depends on the problem to be solved. In our study, precision denotes the ratio of correct prediction among all the images predicted as hit, while recall denotes the ratio of correct prediction among all the images annotated as hit. Images with diffraction signals should be identified as correctly as possible, thus the model should achieve a high recall and while a relatively low precision is acceptable.

\subsection{Benchmark}
Firstly, a two-dimensional (2D) CNN are trained with sufficient labeled data as a benchmark. For each dataset, the total 2000 images are divided into 1200, 400, 400 for training, validation, and test, respectively. The architecture of the CNN is shown in Fig. \ref{fig-typical-cnn}, and the results are summarized in Table \ref{tab-2D-CNN}. For comparison, each trained 2D CNN is directly tested on the other two datasets. It can be seen that a 2D CNN trained with sufficient labeled images performs very well on its own dataset, but not so well on other datasets with many ``Hit or Maybe'' images being misclassified.
\begin{table}[ht!]
	\centering
	\begin{threeparttable}
		\caption{\label{tab-2D-CNN}The accuracy of 2D CNN with sufficient training.}
		\begin{tabular}{>{\centering}m{35pt} >{\centering}m{55pt}  >{\centering}m{30pt}  >{\centering}m{55pt}  >{\centering}m{30pt} >{\centering}m{55pt} >{\centering\arraybackslash}m{30pt}} 
			\toprule
			& \multicolumn{6}{c}{Prediction} \\
			\cmidrule(lr){2-7}
			\multirow{2}[3]{*}{} & \multicolumn{2}{c}{LN83} & \multicolumn{2}{c}{LN84} & \multicolumn{2}{c}{LO19} \\
			\cmidrule(lr){2-3} \cmidrule(lr){4-5} \cmidrule(lr){6-7}
			Training & Hit or Maybe & Miss & Hit or Maybe & Miss & Hit or Maybe & Miss \\
			\midrule
			LN83 & \textbf{96.92\%}  &  99.40\%  & \textbf{90.28\% } &  98.91\%  & \textbf{70.37\%}  & 99.58\% \\
			\cmidrule{1-7}
			LN84 & \textbf{67.69\%} & 100.00\%  & \textbf{94.91\% } & 97.83\%  & \textbf{81.48\% }   & 100.00\% \\
			\cmidrule{1-7}
			LO19 & \textbf{95.38\%} &99.40\%    & \textbf{87.50\% } & 98.91\% & \textbf{96.30\% }   & 97.06\% \\
			\bottomrule
		\end{tabular}
	\end{threeparttable}
\end{table}

For each dataset, we also build an insufficiently trained 2D CNN model by dividing the total 2000 images into training, validating and test sets with 200, 200, 1600 images, respectively. The prediction accuracies of those models are shown in Table \ref{tab-2D-CNN-Insufficient}. Compared with Table \ref{tab-2D-CNN}, it can be seen that the recalls are much worse, namely many ``Hit or Maybe'' patterns are misclassified as ``Miss''. The poor recalls indicate that the 2D CNN trained by a small labeled dataset is unable to learn the features of the ``Hit or Maybe'' frames.
\begin{table}[ht!]
	\centering
	\begin{threeparttable}
		\caption{\label{tab-2D-CNN-Insufficient}The accuracy of 2D CNN with insufficient training.}
		\begin{tabular}{>{\centering}m{30pt} >{\centering}m{60pt} >{\centering\arraybackslash}m{30pt}}
			\toprule
			Datasets & Hit or Maybe & Miss  \\
			\midrule
			LN83 & \textbf{51.06\%}  &  100.00\%  \\
			\cmidrule{1-3}
			LN84 & \textbf{65.07\%} &  99.76\%  \\
			\cmidrule{1-3}
			LO19 & \textbf{84.87\%}  &  99.56\%  \\
			\bottomrule
		\end{tabular}
	\end{threeparttable}
\end{table}

\subsection{Results of weakly supervised models}
\subsubsection{Transfer learning}
Following the steps depicted in Fig. \ref{fig-TF-learning}, the new CNNs are constructed by reusing the convolutional bases of the fully trained 2D CNNs and then fine-tuned by 200 labeled frames from the other two datasets. Each transferred model is validated with 200 images and tested with 1600 images. The results are given in Table \ref{tab-TF}. Compared with the predictions in Table \ref{tab-2D-CNN}, it can be seen that the recalls have been improved significantly after the fine tuning. Furthermore, compared with training a 2D CNN from scratch by the same 200 labeled images (see Table \ref{tab-2D-CNN-Insufficient}), it can be concluded that reusing the convolutional base of a well-trained 2D CNN is a much better approach. It is important to note that those three datasets were acquired with the same instrument and detector, but different photon energies, samples and sample delivery devices \cite{Ke2018}. It seems that the SFX datasets acquired with the same instrument and detector have some hidden similarities and thus transfer learning works well.
\begin{table}[ht!]
	\centering
	\begin{threeparttable}
		\caption{\label{tab-TF} The accuracy of transfer learning.}
			\begin{tabular}{>{\centering}m{35pt} >{\centering}m{55pt}  >{\centering}m{30pt}  >{\centering}m{55pt}  >{\centering}m{30pt} >{\centering}m{55pt} >{\centering\arraybackslash}m{30pt}} 
			\toprule
			& \multicolumn{6}{c}{Prediction} \\
			\cmidrule(lr){2-7}
			\multirow{2}[3]{*}{} & \multicolumn{2}{c}{LN83} & \multicolumn{2}{c}{LN84} & \multicolumn{2}{c}{LO19} \\
			\cmidrule(lr){2-3} \cmidrule(lr){4-5} \cmidrule(lr){6-7}
			Training & Hit or Maybe & Miss & Hit or Maybe & Miss & Hit or Maybe & Miss \\
			\midrule
			LN83 & -  &  -  & \textbf{96.68\%}  &  92.33\%  & \textbf{95.82\%}  & 93.38\% \\
			\cmidrule{1-7}
			LN84 & \textbf{94.53\%} & 91.82\%  & - & -  & \textbf{94.24\%}    & 93.49\% \\
			\cmidrule{1-7}
			LO19 & \textbf{98.18\%} &97.40\%    & \textbf{98.01\%}  & 91.62\% & -  & - \\
			\bottomrule
		\end{tabular}
	\end{threeparttable}
\end{table}

\subsubsection{Dimensionality Reduction}\label{sec-row-wise}
\begin{figure}[ht!]
	\centering
	\includegraphics[width=1.\textwidth]{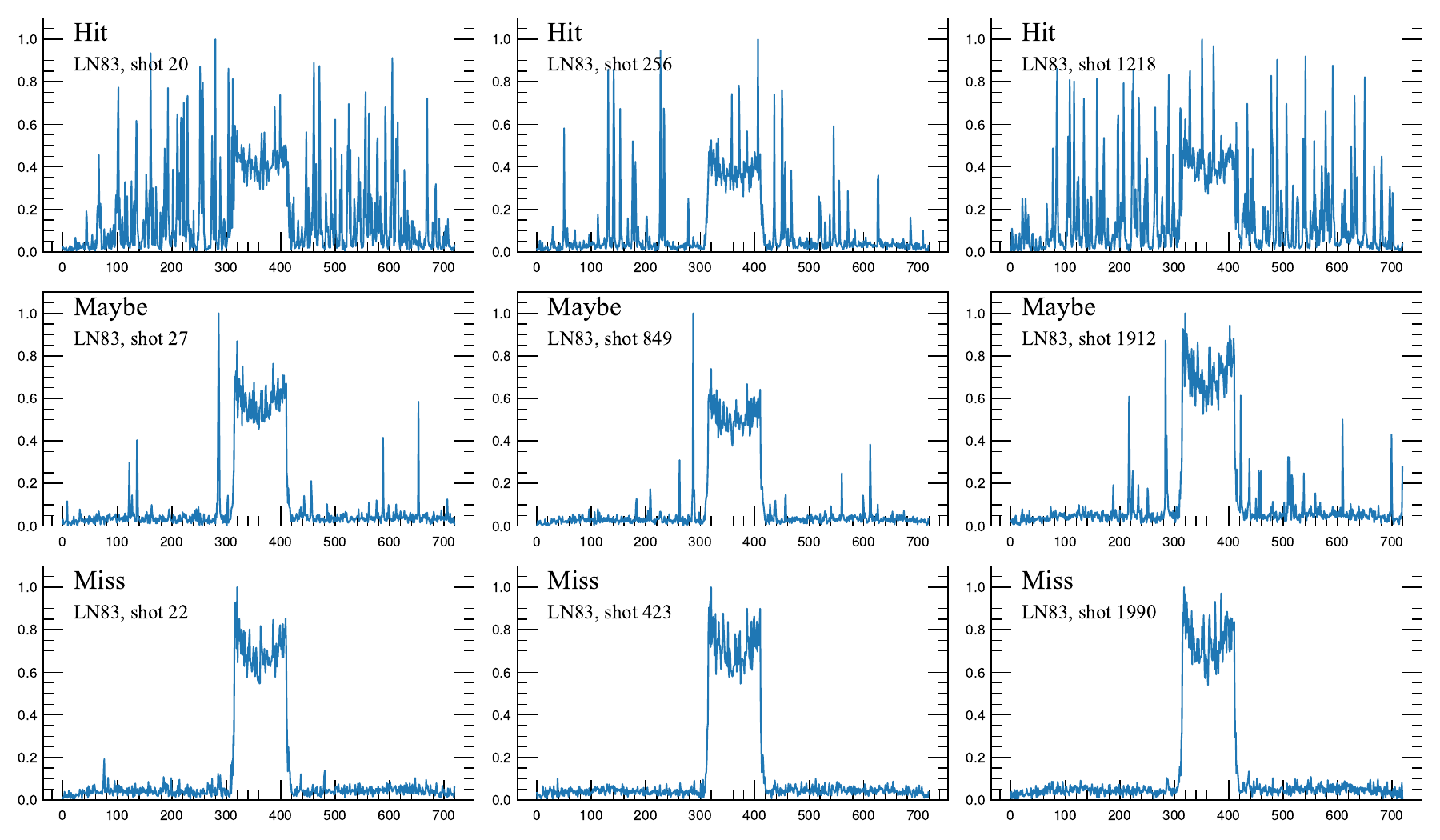}
	\caption{The 1D vectors produced by row-wise decomposition clearly show the distinguishing features between ``Miss'', ``Hit'' and ``Maybe''.}
	\label{fig-row-wise-decom}
\end{figure}

The essential difference between ``Miss'' and ``Hit or Maybe'' images is that the latter has more Bragg spots, which are characterized by high photon counts. Furthermore, Bragg spots are often sparsely distributed over the whole image, suggesting that the image may contain some information that is redundant for the purpose of classification. These two facts inspire us to transform the diffraction patterns from 2D space to 1D space. For each image, we first scan the image to find the pixels with maximum and minimum values in each row, and then perform the following subtraction:
\begin{equation}\label{eq-decom}
d_{i} = p_{i}^{\mathrm{max}} - p_{i}^{\mathrm{min}},
\end{equation}
where $i = 1, 2, 3, ..., 720$ is the row index, $p_{i}^{\mathrm{max}}$ and $p_{i}^{\mathrm{min}}$ denote the maximum and minimum values in the $i$-th row, respectively. Consequently, in each dataset all images are converted to 2000 1D vectors: $\mathbf{D}_{1}, \mathbf{D}_{2}, \cdots, \mathbf{D}_{2000}$, where $\mathbf{D}_{j} = (d_{1}, d_{2}, \cdots, d_{720})$. This transformation is called row-wise decomposition (RWD) in our work. The 1D vectors clearly show the distinguishing features between ``Miss'', ``Hit'' and ``Maybe'' (Fig. \ref{fig-row-wise-decom})

The 2000 1D vectors are divided into training set, validation set and test set with 200, 200, 1600 samples, respectively. Then a 1D CNN is trained and tested by those new sets. The predictions of all datasets are shown in Table \ref{tab-1D-CNN}, from which it can be seen that 200 labeled samples are able to generate models with good performance after the dimensionality reduction, while only weak CNN models can be produced before that (see  Table \ref{tab-2D-CNN-Insufficient}).
Additionally, 1D CNN processes 11,123 vectors per second in the test, while the 2D CNN can only process 126 frames per second on the same device (A NVIDIA Tesla A100 GPU card with the memory of 40 GByte). Therefore dimensionality reduction is a very promising approach to screening the diffraction frames in real-time.
\begin{table}[ht!]
	\centering
	\begin{threeparttable}
		\caption{\label{tab-1D-CNN}The accuracy of 1D CNN after dimensionality reduction.}
		\begin{tabular}{>{\centering}m{50pt} >{\centering}m{80pt}  >{\centering\arraybackslash}m{50pt}}
			\toprule
			Datasets & Hit or Maybe & Miss  \\
			\midrule
			LN83 & \textbf{95.14\%}  &  96.22\%   \\
			\cmidrule{1-3}
			LN84 & \textbf{94.95\%} &  86.19\%   \\
			\cmidrule{1-3}
			LO19 & \textbf{95.24\%}  &  94.04\%  \\
			\bottomrule
		\end{tabular}
	\end{threeparttable}
\end{table}

\subsubsection{DANN}
As described in  Section \ref{sec:DANN}, the training of a DANN model needs two datasets. We compose six pairs of datasets, one for source and another for target (labels are ignored in training), and then train the DANNs. At the end of training, the convolutional base is able to extract features from the common characteristic spaces of the source and target, meanwhile the label classifier is able to identify the extracted features as ``Hit or Maybe'' or ``Miss''. Finally, the CNN composed of the convolutional base and the label classifier is tested on the entire target dataset with human annotations. The results for all models are shown in Table \ref{tab-DANN}.
\begin{table}[ht!]
	\centering
	\begin{threeparttable}
		\caption{\label{tab-DANN}The accuracy of DANN models.}
		\begin{tabular}{>{\centering}m{24pt} >{\centering}m{24pt}  >{\centering}m{52pt}  >{\centering}m{24pt}  >{\centering}m{52pt}  >{\centering}m{24pt}  >{\centering}m{52pt}  >{\centering\arraybackslash}m{24pt}} 
			\toprule
			& &  \multicolumn{6}{c}{Prediction} \\
			\cmidrule(lr){3-8}
			\multirow{2}[3]{*}{} & &  \multicolumn{2}{c}{LN83} & \multicolumn{2}{c}{LN84} & \multicolumn{2}{c}{LO19} \\
			\cmidrule(lr){3-4} \cmidrule(lr){5-6} \cmidrule(lr){7-8}
			Source & Target & Hit or Maybe & Miss & Hit or Maybe & Miss & Hit or Maybe & Miss \\
			\midrule
			\multirow{2}[3]{*}{LN83} & LN84 & \textbf{90.77\%} & 98.81\% & \textbf{90.19\%} & 91.9\% & - & -  \\
			\cmidrule{2-8}
			& LO19 & \textbf{92.31\%} & 98.81\% & - & - & \textbf{89.67\%} & 98.39\% \\
			\cmidrule{1-8}
			\multirow{2}[3]{*}{LN84} & LN83 & \textbf{92.18\%} & 91.77\% & \textbf{92.59\%} & 97.28\% & - & - \\
			\cmidrule{2-8}
			& LO19  & - & - & \textbf{91.67\%} & 97.28\% & \textbf{85.58\%} & 96.43\%  \\
			\cmidrule{1-8}
			\multirow{2}[3]{*}{LO19} & LN83 & \textbf{96.09\%} & 97.36\% & - & - & \textbf{92.59\%} & 97.90\% \\
			\cmidrule{2-8}
			& LN84 & -  & -  & \textbf{90.41\%} & 91.53\%  & \textbf{86.42\%} & 97.00\%  \\
			\bottomrule
		\end{tabular}
	\end{threeparttable}
\end{table}

\begin{figure}[ht!]
	\centering
	\includegraphics[width=0.9\textwidth]{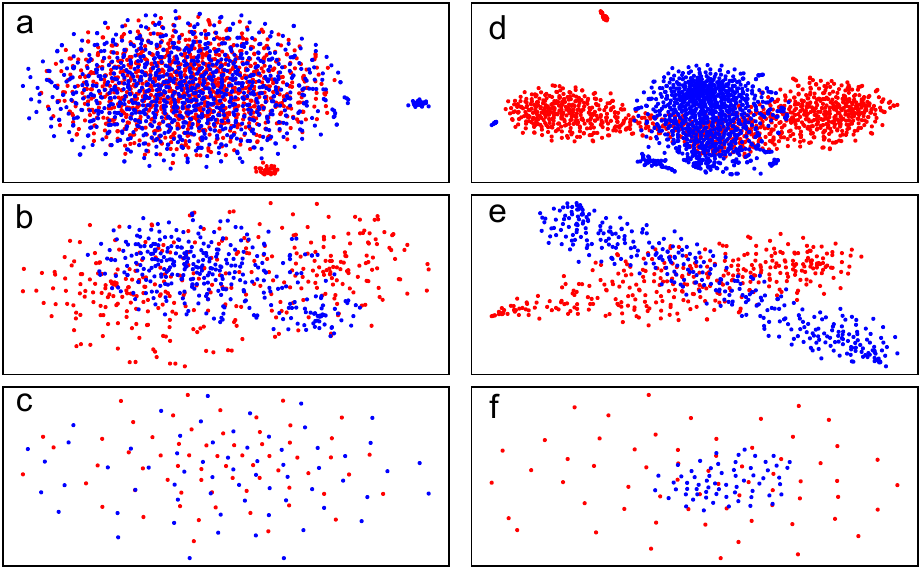}
	\caption{Comparison of the features learned by the DANN and the CNN trained only on source. The extracted features by deep learning models are usually high-dimensional, hence all the features are embedded into 2D space through t-SNE algorithm for visualization. Red and blue dots represent 2D embeddings of features extracted from the source (LN83) and target(LO19), respectively, and the more overlap between their distributions indicates that the model is more capable of learning features from the common space of these two datasets. Models and labels of the features shown in each panel are: (a) DANN, miss; (b) DANN, maybe; (c) DANN, hit; (d) CNN, miss; (e) CNN, maybe; (f) CNN, hit.}
	\label{fig-compare-feature}
\end{figure}

The essential premise of DANN is that the source and target datasets share a common characteristic space. Because the main features of ``Hit'' and ``Miss'' frames are clear and simple, i.e., ten or more Bragg spots for ``Hit'' and three or less Bragg spots for ``Miss'', it is reasonable to assume that ``Hit'' and ``Miss'' frames from both datasets have similar latent features. However, the features of ``Maybe'' frames can be different.
In order to visualize the high-dimensional features learned by the DANN model, the extracted features, i.e., the output of convolutional base (see Fig. \ref{fig-DANN-architecture}), are projected into 2D space through t-distributed stochastic neighbor embedding (t-SNE) algorithm~\cite{Maaten2008} using the toolkit scikit-learn v1.0.2~\cite{Pedregosa2011}. t-SNE algorithm embeds each high-dimensional instance into low-dimensional space in a way that similar instance are modeled by nearby points while dissimilar instances are modeled by distant points with high probability.
The distributions of the projected features of ``Miss'', ``Maybe'' and ``Hit'' frames of LN83 (source, red dots) and LO19 (target, blue dots) are shown in Fig. \ref{fig-compare-feature} (panels a, b, and c, respectively), and the same distributions produced from the CNN trained on LN83 are also given as a comparison (panels d, e and f, respectively).
The more overlap between the distributions of the two embedded features indicates that the model is more capable of learning features from the common space of these two datasets.
It is clear from Fig. \ref{fig-compare-feature} that DANN is better at learning features from the common spaces than ordinary CNN, which is consistent with the significant improvement of recall on the target dataset (89.67\% vs. 70.37\%, see Table \ref{tab-DANN} and Table \ref{tab-2D-CNN}). On the other hand, there are still some distinctions between the characteristic spaces of source and target, especially for ``Maybe'' frames (See panels b in Fig. \ref{fig-compare-feature}). Probably that is the reason why DANN performs much better on the target than the CNN trained only by the source, but slightly worse than the model fine-tuned by the target.

\section{Discussion}
The three methods above have  their own advantages and disadvantages in classifying diffraction patterns. The comparison in a few aspects is given in Table \ref{tab-comparison}.
The RWD algorithm for dimensionality reduction in our work (See section \ref{sec-row-wise}) looks simple and achieves superior performance, but the potential drawback is the lack of general applicability. In other words, the decomposition in this study may not work in other classification tasks of diffraction patterns, in which case we may need to think about a new method for dimensionality reduction or feature engineering. There may be various feature extraction methods for SFX data, but it is not an easy task to determine which method works the best \cite{Rahmani2023}. That is why we mark the pre-processing of RWD as hard in Table \ref{tab-comparison}.
Nevertheless, RWD algorithm has two important properties in screening diffraction patterns. First, the 1D CNN model based on RWD achieves a speed of over 11 thousands frames per second in labels prediction, which is more than 80 times faster than 2D CNN models (See section \ref{sec-row-wise}). As the pulse frequency of XFELs increases to tens of thousands or even higher \cite{EuXFEL-comparison}, there is a great potential for the RWD algorithm to be developed into an online screening tool for SFX experiments. Secondly, RWD has introduced a new feature that is easier to recognize, and thus appears to be more universal than original features in identifying whether a diffraction images contain physical signals. Without any tuning, we test the RWD-based models and the fully supervised 2D CNN models across datasets, and the result clearly indicates that the former shows a higher level of adaptability and generalization on the datasets collected with different experimental settings (Table \ref{tab-cross-test}). It is desirable to train a universal models which can be used to screen images in various experiments, although it may be difficult \cite{Ke2018, Rahmani2023}. Our work indicates that training a model to learn some one-dimensional features may be a promising approach.
\begin{table}[ht!]
    \centering
    \begin{threeparttable}
        \caption{\label{tab-comparison}Comparison of the three weakly supervised methods.}
        \begin{tabular}{>{\raggedright}m{90pt} >{\raggedright}m{75pt}  >{\raggedright}m{40pt} >{\raggedright\arraybackslash}m{40pt}} 
            \toprule
            & Transfer learning & RWD &  DANN \\
            \midrule
            Size of labels & Low  &  Low & None \\
            Pre-processing & Easy & Hard & Easy \\
            Train & Medium & Easy & Hard \\
            Recall & High & High & Medium \\
            Resource consuming & Medium & Low & High \\
            Speed of prediction & Slow & Fast & Slow \\
            \bottomrule
        \end{tabular}
    \end{threeparttable}
\end{table}

\begin{table}[ht!]
	\centering
	\small
	\begin{threeparttable}
		\caption{\label{tab-cross-test} Cross-dataset test of  RWD-based 1D CNNs and fully supervised 2D CNNs$^{a}$}
			\begin{tabular}{>{\centering}m{22pt} >{\centering}m{35pt}  >{\centering}m{47pt}  >{\centering}m{24pt}  >{\centering}m{47pt}  >{\centering}m{24pt}  >{\centering}m{47pt}  >{\centering\arraybackslash}m{24pt}} 
				\toprule
				& &  \multicolumn{6}{c}{Prediction} \\
				\cmidrule(lr){3-8}
				\multirow{2}[3]{*}{} & &  \multicolumn{2}{c}{LN83} & \multicolumn{2}{c}{LN84} & \multicolumn{2}{c}{LO19} \\
				\cmidrule(lr){3-4} \cmidrule(lr){5-6} \cmidrule(lr){7-8}
				Training & Model & Hit or Maybe & Miss & Hit or Maybe & Miss & Hit or Maybe & Miss \\
				\midrule
				\multirow{2}[3]{*}{LN83} & 1D CNN & - & - & \textbf{96.91\%} & 78.53\% & \textbf{90.01\%} & 98.39\%  \\
				\cmidrule{2-8}
				& 2D CNN & - & - & \textbf{90.28\% } & 98.91\% & \textbf{70.37\%} & 99.58\% \\
				\cmidrule{1-8}
				\multirow{2}[3]{*}{LN84} & 1D CNN & \textbf{94.87\%} & 95.79\% & - & - & \textbf{91.94\%} & 96.60\% \\
				\cmidrule{2-8}
				& 2D CNN  & \textbf{67.69\%} & 100.00\% & - & - & \textbf{81.48\%} & 100.00\%  \\
				\cmidrule{1-8}
				\multirow{2}[3]{*}{LO19} & 1D CNN & \textbf{97.07\%} & 94.03\% & \textbf{96.27\%} & 86.06\% & - & - \\
				\cmidrule{2-8}
				& 2D CNN & \textbf{95.38\%}  & 99.40\% & \textbf{87.50\%} & 98.91\%  & - & -  \\
				\bottomrule
			\end{tabular}
			\centering{$^{a}$2D CNNs in this table are the models trained with 1200 labeled patterns.}
		\end{threeparttable}
	\end{table}

\begin{figure}[ht!]
	\centering
	\includegraphics[width=0.95\textwidth]{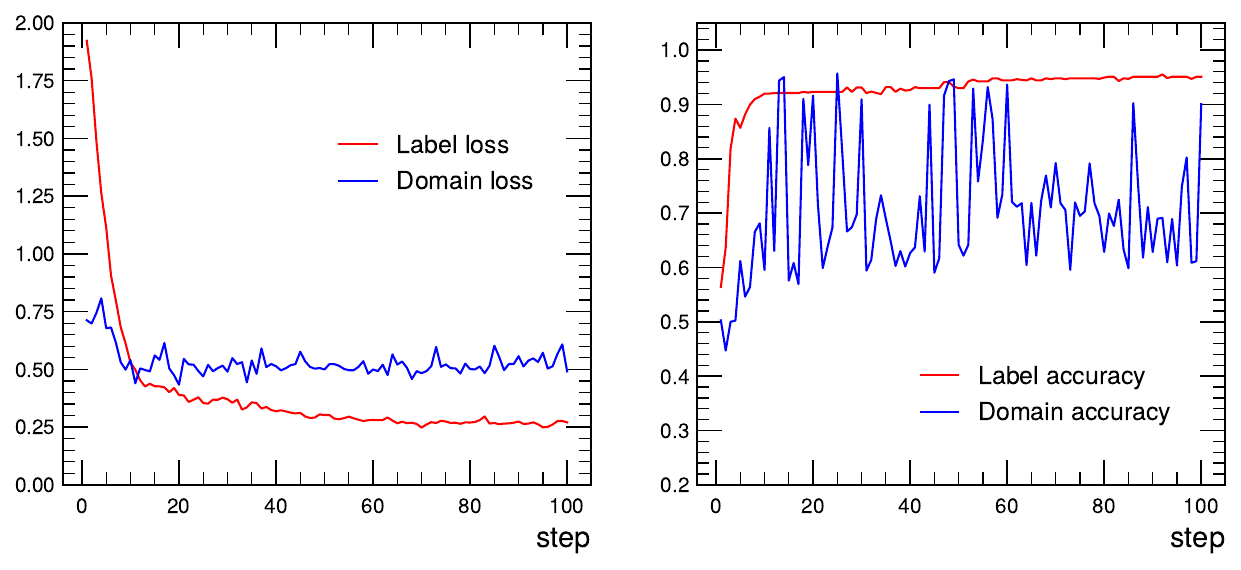}
	\caption{In the validation steps of a DANN model, the loss and accuracy of label prediction converge to plateaus gradually, while those two metrics of domain prediction tend to oscillate.}
	\label{fig-domain-oscillation}
\end{figure}

Both transfer learning and DANN rely on another annotated dataset acquired in similar experimental setup, but their training processes are quite different. The domain classifier of a DANN model is designed to be fooled by the domains of source or target, thus its loss and accuracy are supposed to oscillate around certain values. In order to verify this hypothesis, we monitor the loss and accuracy in validation steps and draw their trends in Fig. \ref{fig-domain-oscillation}, from which it can be seen that the loss and accuracy of the label prediction converge to plateaus gradually, whereas those two metrics of domain prediction tend to oscillate. It is reasonable to assume that the two plateaus and two oscillations indicate that the DANN model has been trained, and thus the labels of the target are no longer needed for validation. Although the transferred models work slightly better, they need extra labeled patterns for fine tuning and validation.

Furthermore, the results of our study indicate that the same instrument and the same detector are able to yield similar latent features in the SFX diffraction frames so that both transfer learning and DANN work well. Therefore, for experimental datasets produced at the same SFX instrument and by the same detector, regardless of the samples, we may only need to fully annotate one dataset, while a small number of annotated frames in other datasets is enough to train a good CNN model.


\section{Conclusion}\label{sec:summary}
In this paper, we studied three weakly supervised models, i.e., transfer learning, dimensionality reduction, and DANN, to classify SFX diffraction patterns and showed that although these models were trained by only 200 labeled samples or even less, they demonstrated comparable performance to the fully supervised CNN models (i.e., trained by 1200 labeled patterns). With the development of advanced X-ray sources, the processing of experimental data is facing increasing challenges and machine learning methods are expected to play an important role. This study demonstrates that the label dependence of CNNs can be greatly reduced, thus providing a promising approach to efficient SFX data processing.

Besides the three methods described above, we are also studying some other algorithms, e.g., unsupervised pre-training by generative adversarial network (GAN)~\cite{Goodfellow2014}, building an ensemble model by combining several weak neural networks \cite{Zhou2009}. In the future we will continue to work on weakly supervised algorithms, with the ultimate goal of finding weakly supervised solutions to various classification tasks, including not only the recognition of ``Hit'' or ``Miss'' images in SFX experiment, but also some problems in similar experiments at XFEL, e.g., selecting single hit frames in single-particle imaging (SPI) data \cite{Li2020,Ekeberg2016,Shi2019}.
There are two XFEL facilities in Shanghai, of which one is Shanghai soft X-ray free-electron laser (SXFEL) ~\cite{Zhao2017,Liu2021} and the other is Shanghai High repetition rate XFEL and Extreme Light Facility (SHINE). SXFEL started its commissioning in 2021 and opened to users in 2022 \cite{Fan2022}. The repetition rate of SHINE is much higher than that of SXFEL, and therefore the data processing of SHINE will also be far more challenging in the future. All the weakly supervised models in our study will be important tools for efficient data analysis in SXFEL and SHINE.

All of the code, written in Python 3.8, using TensorFlow framework 2.3 and executable in JupyterLab 3.2, is available on the request by email.

\begin{backmatter}
\bmsection{Funding}
This work is financially supported by Strategic Priority Research Program of Chinese Academy of Sciences (Grant No. XDC02070100).

\bmsection{Acknowledgments}
We acknowledge Ke \textit{et al.} for the deposition of datasets on CXIDB for open access. We also acknowledge the help provided by the course "CS286: AI for Science and Engineering" at ShanghaiTech University. Finally we would like to thank Yaru Yin and Wujun Shi for fruitful discussions.

\bmsection{Disclosures}
The authors declare no conflicts of interest.

\bmsection{Data availability}
Data underlying the results presented in this paper are available in Ref. \cite{Ke2018}.
\end{backmatter}



\end{document}